\documentclass[journal=jpccck,layout=twocolumn]{achemso}
\usepackage[version=3]{mhchem} 
\usepackage[T1]{fontenc}       
\usepackage{color}

\author{Qian Li}
\affiliation[]{Nano-Science Center and Department of Chemistry, University of Copenhagen, DK-2100 Copenhagen {\O}, Denmark}

\author{Ivan Duchemin}
\affiliation[]{ Univ Grenoble Alpes, F-38000 Grenoble, France}
\affiliation[]{ CEA, INAC SP2M/L$\_$Sim, F-38054 Grenoble, France}

\author{Shiyun Xiong}
\affiliation[]{Max Planck Institute for Polymer Research, Ackermannweg 10, 55128 Mainz, Germany}

\author{Gemma C. Solomon}
\affiliation[]{Nano-Science Center and Department of Chemistry, University of Copenhagen, DK-2100 Copenhagen {\O}, Denmark}

\author{Davide Donadio}
\email{ddonadio@ucdavis.edu}
\affiliation[]{Department of Chemistry, University of California Davis, One Shields Avenue, Davis, CA, 95616}
\affiliation[]{Max Planck Institute for Polymer Research, Ackermannweg 10, 55128 Mainz, Germany}
\affiliation[]{DIPC, Donostia International Physics Center, Paseo Manuel del Lardizabal 4, 20018, San Sebastian, Spain}
\affiliation[]{Ikerbasque, Basque Foundation for Science, Bilbao, Spain}

\title[An \textsf{achemso} demo]
  {Mechanical Tuning of Thermal Transport in a Molecular Junction} 

\abbreviations{IR,NMR,UV}

\begin{document}

\begin{tocentry}

 \includegraphics[width=1.0\linewidth]{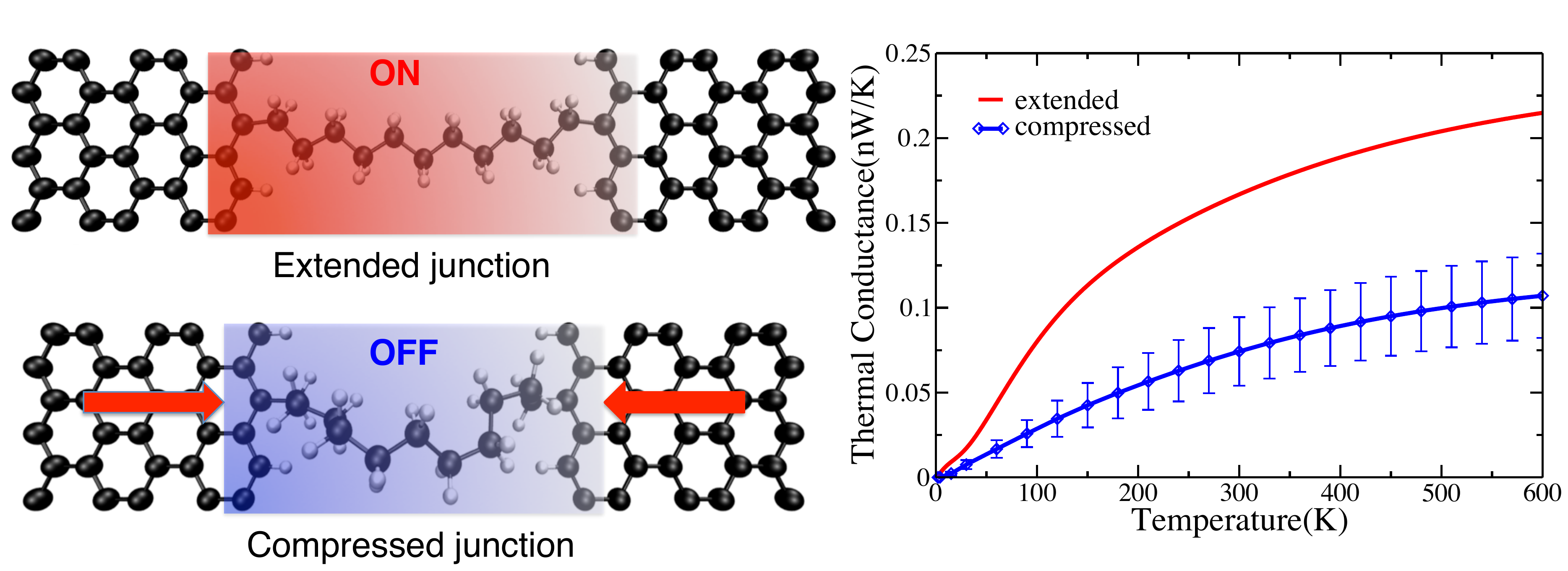}
  
\end{tocentry}

\begin{abstract}
  Understanding and controlling heat transport in molecular junctions would provide new routes to design nanoscale coupled electronic and  phononic devices.~\textcolor{black}{Using first principles full quantum calculations, we tune thermal conductance of a molecular junction by mechanically compressing and extending a short alkane chain connected to graphene leads.} We find that the thermal conductance of the compressed junction drops by half in comparison to the extended junction, making it possible to turn on and off the heat current. The low conductance of the off state does not vary by further approaching the leads and stems from the suppression of the transmission of the in--plane transverse and longitudinal channels. \textcolor{black}{Furthermore, we show that misalignment of the leads does not reduce the conductance ratio}. These results also contribute to the general understanding of thermal transport in molecular junctions.
\end{abstract}


\section{Introduction}

Thermal transport at the level of an individual molecule cannot be described using standard concepts from macroscopic theories~\cite{Dubi2011}. While 10 years of research in molecular electronics have led to significant developments in our understanding of charge transport, the question of how heat transport is influenced by the structure and the chemistry of molecular junctions is still largely unexplored. 
The experimental techniques commonly used in molecular electronics make it possible to readily apply thermal and mechanical perturbations at the molecular scale: for example, a single molecule can be stretched or compressed using either scanning tunnel microscopy (STM) or  mechanically controlled break-junction (MCBJ) techniques~\cite{Jacs2008, Tao2011, nature2015}. Electronic conductance and thermopower show large variations depending on different conformations of molecules that can be determined by strain~\cite{Jacs2008,Su:2015jt,Vacek:2015js}. In turn, it is still unclear how physical perturbations influence phononic transport in molecular junctions, and whether it is possible to exploit the interplay between the thermal and mechanical properties of the molecule to tune heat transport. A better control of thermal transport in single molecule devices and self assembled monolayers is crucial for designing molecular electronic circuits with efficient heat dissipation and for realizing high performance thermoelectric junctions~\cite{Dresselhaus2007, Reddy-science, Bergfield2010}. In addition molecular junctions provide a versatile platform to design nanophononic devices, such as thermal switches and rectifier~\cite{Baowen2012, Martin2013}.

Progresses in experimental techniques enabled measurements of the thermal conductance across heterointerfaces between solids and monolayers of organic molecules, and heat dissipation in molecular junctions~\cite{Reddy2013, science2007, Losego2012}. Recently Meier et al. reported the length dependence of thermal transport across self-assembled monolayers of alkane chains using scanning thermal microscopy~\cite{Gotsman2014}. 
A few theoretical studies have considered phonon transport in molecular junctions using molecular dynamics (MD) or lattice dynamics with empirical interatomic potentials, identifying some general trends~\cite{Nitzan, HuLin2010, Keblinski2011, Duda2011,Velizhanin:2011cl}.  Markussen studied phonon interference effects in cross-conjugated molecular junctions~\cite{Markussen2013}, and L\"u compared the thermal conductance of  benzene and alkane chains clamped between carbon nanotube leads~\cite{Jingtao2008} using Density functional theory (DFT) combined with Green's function within the harmonic approximation. The same approach was also used to calculate lattice thermal conductance in thermoelectric molecular junctions~\cite{Cuniberti2010, Markussen2012}. However, further efforts on the theoretical side are necessary to achieve a comprehensive picture of such a complex phenomenon as phonon transport in molecular junction.
 
Polyethylene chains of different lengths are chemically simple, yet very interesting systems to investigate heat transport at the molecular scale. Single Polyethylene chains were predicted to exhibit very high thermal conductivity by MD simulations, eventually confirmed in experiments on drawn fibers~\cite{Henry2008kg, Shen2010fr}. Zhang {\it et al.},  using MD simulations, proposed thermal conductivity regulation for polyethylene nanofibers as a result of a transition between a disordered phase and a phase in which there are domains of aligned polymer chains~\cite{Zhang2013}.  
Sasikumar et al.~\cite{Keblinski2011} computed phonon transport through polyethylene chains bonded to silicon leads by non-equilibrium MD and predicted a two-fold increase of the thermal conductance upon extension of the polymer chain, when all monomers are in  $\it trans$ configuration. 
However insightful, this study leaves several open questions, especially concerning the extension of the results to low temperatures, deep in the phonon quantum regime, since MD lacks a proper treatment of quantum effects. A parameter-free first-principles method in which both electrons and phonons are correctly treated as quantum entities is necessary for a detailed quantitative study of phonon transport in molecular junctions~\cite{Mingo2008}.
Furthermore, it is worth investigating the effect of leads with a broader range of frequencies, matching the modes of polyethylene, and probing the specific contribution of phonons with different polarization to heat transport.  

In this paper we study the thermal conductance of a molecular junction, consisting of a polyethylene chain bonded to graphene leads, as a function of the distance between the leads and of their alignment. 
\textcolor{black}{Graphene-molecule-graphene junctions, which are ultraflat, transparent, robust, exhibit stable electrical characteristics~\cite{Ullmann2015, Prins2011}. For thermal properties, graphene has the unique nature of two-dimensional phonon transport, which has recently attracted significant attention~\cite{Balandin2008, Balandin2011, Nika2012}.
Vibrations in the graphene lattice are characterized by two types of phonons: those vibrating in the plane of the layer and those with vibrations out of the plane of the layer (flexural phonons)~\cite{Lindsay2010}.
Compared with the metal leads, graphene has a broader range of frequencies, which can reduce the effect of vibrational mismatch with the molecule. Thus, we employ graphene leads, covalently bound to our molecule, to probe the question of how the injected vibrational energy is scattered by a single molecule.} 


We employ first-principles DFT to optimize the structure and to calculate the force constant matrix, which is eventually used to compute the thermal conductance of the molecular junction {\it via} the elastic scattering matrix approach~\cite{Young1989tm}. The structural complexity of the compressed junction, which exhibits a large number of nearly iso-energetic minima, is sampled by MD, and the thermal conductance is computed for representative "inherent" structures~\cite{Stillinger1983}, identified by cluster analysis~\cite{Daura1999}. 
We first discuss the reduction of thermal conductance upon compression of the junction as a function of temperature, and we analyze it in terms of phonon transmission by polarization. \textcolor{black}{Our results confirm that the thermal conductance of the polyethylene junction can be modulated mechanically, and we verify that the switching mechanism remains effective over a broad range of temperatures and also with misalignment of the graphene leads}.

\section{Systems and methods}

\begin{figure}[H]
\centering
  \includegraphics[width=0.85\linewidth]{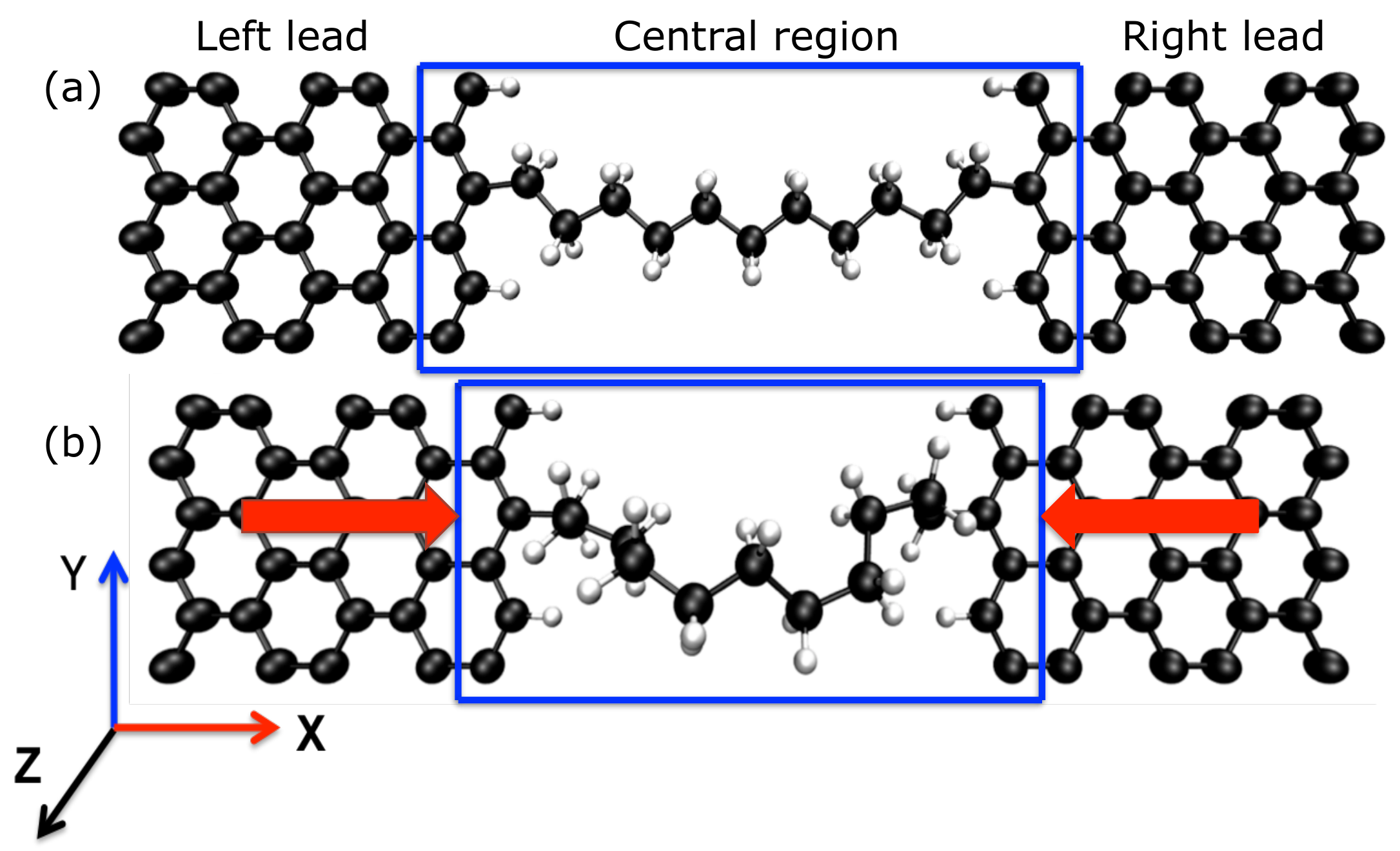}
  \caption{The molecular junction with an extended (a) and compressed (b) alkane chain made of 11 units . The blue rectangle highlights the device region in the elastic scattering calculations.}
  \label{junction}
\end{figure}

We consider a molecular junction made of an alkane chain with 11 (CH$_2$) groups clamped between two semi-infinite periodic graphene leads, as shown in Fig.~\ref{junction}. In the elastic phonon scattering calculation the system is divided into three parts: the central region and two leads acting as thermal baths at different temperature. The device is oriented so that heat flows along the $x$ axis. The central region includes a part of the graphene leads, which are saturated with hydrogen atoms. 
We fix the outer graphene leads and relax the central region using DFT. Then the force constant matrix is calculated from the relaxed structure. Periodic boundary conditions are applied in the calculations of the force constant matrices of the central region and of the graphene leads. Periodic replicas of the system in the $z$ direction are separated by a vacuum region of 10 \AA, which is sufficient to make their interactions negligible.

We computed phonon transport using lattice dynamics and the elastic scattering kernel method~\cite{Ivan2011}. 
The use of the harmonic approximation is justified by the former observation that anharmonic scattering is of limited importance for short all-carbon molecular junctions at room temperature and below \cite{Nitzan,Keblinski2011}. 
In harmonic lattice dynamics the phonon properties are fully determined by the force constant matrix, $K$, which is calculated by finite differences:
\begin{equation}
  K_{i\alpha, j\beta} =  \frac{\partial^2 E}{\partial\mu_{i\alpha} \partial\mu_{j\beta}} =  \frac{F_{j\beta}(Q_{i\alpha}) - F_{j\beta}(-Q_{i\alpha})}{2Q_{i\alpha}}
\end{equation}
where $\mu_{i\alpha}$($\mu_{j\beta}$) is the displacement of atom $i$($j$) in the $\alpha$($\beta$) coordinate direction and $E$ is the total energy. The central region of molecular junctions is initially relaxed to a maximum residual force of 0.001 Ry/Bohr. Then each atom $i$, is displaced by $Q_{i\alpha} =\pm0.02$ \AA~in the direction $\alpha = {x, y, z}$ to obtain the forces $F_{j\beta}(Q_{i\alpha}$), on atom $j$ along the $\beta$ coordinate.
From the force constant matrix, we get the dynamical matrix $D$, with $D_{i\alpha, j\beta}  =  K_{i\alpha, j\beta}/\sqrt{m_i m_j} $, where $m_i$($m_j$) is the mass of atom $i(j)$.

The eigen-frequencies, $\omega_i$ and eigen-vectors, $\mu_i$ can be obtained by solving the linear system  $D \epsilon_i = {\omega_i}^2\epsilon_i$.
The dynamical matrices of the periodic leads and of the device are used to compute the elastic scattering matrix ~\cite{Young1989tm} using the kernel method~\cite{Ivan2011}. The transmission function between the two thermal baths A and B is expressed as:
\begin{equation}
 \tau_{AB}(\omega) = \sum_{i\in A} \sum_{j\in B} |S_{i j}(\omega)|^2
\end{equation}
where the scattering tensor $S(\omega)$ maps for each frequency $\omega$ the incoming phonons onto the outgoing phonons. 
The corresponding thermal conductance is given by the Laudauer formula:
\begin{equation}
  G_{AB}(T) = \int^{\infty}_{0} \frac{\hbar \omega}{2\pi} \tau_{AB}(\omega) \frac{\partial f(\omega, T)}{\partial T} d\omega
\end{equation}
where $f(\omega, T) = \frac{1}{e^{\hbar \omega/{k_B T}} - 1}$ is the Bose-Einstein distribution function at the reservoir temperature T, in which we assume that temperature difference between the two thermal baths tends to zero. 
Using this approach with the setup depicted in Fig.~\ref{junction} one treats ripples in graphene leads as out-of-plane phonons in linear response. Possible non-linear effects related to large ripples~\cite{Fasolino:2007}, mostly occurring at high temperature, would not be taken into account. 

Geometry optimizations and the calculation of force constant matrices are performed using DFT as implemented in the software package Quantum Espresso~\cite{QE}. We use the generalized gradient corrected exchange and correlation PBE functional~\cite{Perdew:1996ug}, and we treat core electrons using ultrasoft pseudopotentials~\cite{PhysRevB.41.7892},. The Kohn-Sham wave functions and the charge density are expanded on plane wave basis sets with energy cutoff of 35 Ry and 300 Ry, respectively. Methfessel-Paxton smearing~\cite{Methfessel1989zz} with an energy width of 0.02 Ry is used. 
We use a 4$\times$4$\times$1 k-point mesh for the graphene leads and a 1$\times$4$\times$1 k-point mesh for the central junction in the phonon calculation to obtain the dynamical matrices. 
Convergence tests on the basis set cutoff, the number of k-points and the size of the vacuum layer in the $z$ direction show that with our settings the phonon frequencies are accurate within 2 $cm^{-1}$.

We first consider a molecular junction with an extended (CH$_2$)$_{11}$  alkane chain, with all the units in {\sl trans} conformation, as shown in Fig.~\ref{junction}. In these conditions the system has a single configuration and does not require statistical sampling.
Eventually we introduce gauche defects into this alkane chain to simulate the compression process. The compressed chain is covalently connected to the graphene leads.  In order to probe thermal conductance as a function of compression, we investigated molecular junctions with different compression ratios ($1-L_{compressed} / L_{extended}$): 0.11,  0.22, corresponding to shifting the leads closer by 1.4 \AA~and 2.8 \AA, respectively.
 \textcolor{black}{These two compression ratios are chosen by introducing a different number of gauche defects. The initial compressed structures, created by the introduction of one or two gauche defects, are just one of many possible configurations at this compression ratio, so we use classical MD at room temperature to sample the configurations of the alkane chain, keeping the graphene leads fixed.}
 
In these MD simulations interatomic interactions are modeled using the Tersoff potential~\cite{tersoff1989}. For each compressed junction, we extracted  50 configurations from the MD trajectory saving frames every  400 fs, and we optimized their geometry using DFT forces, so to obtain zero temperature inherent structures~\cite{Stillinger1983}. 
Since the calculations of the force constants for so many conformations would be too computationally expensive, we perform a cluster analysis to identify a subset of independent structures.  
The cluster analysis is carried out using the algorithm described in Ref.\cite{Daura1999} based on the root mean square differences (RMSD) in atomic positions. RMSD cutoffs of 0.62 and 0.69 \AA, for the 1.4 \AA\ and 2.8 \AA\ compressed junctions respectively, were used to define neighboring structures. This procedure yields 9 independent structures for the 1.4 \AA\ compressed junction and 10 for the 2.8 \AA.
The force constants  are calculated for the representative structure of each cluster by DFT. Finally, the phonon transmission function and conductance are calculated as a weighted average over the representative structures, using the number of elements in each cluster as a weighting factor.

\section{Results and discussion}

\begin{figure}[H]
\centering
  \includegraphics[width=1.0\linewidth]{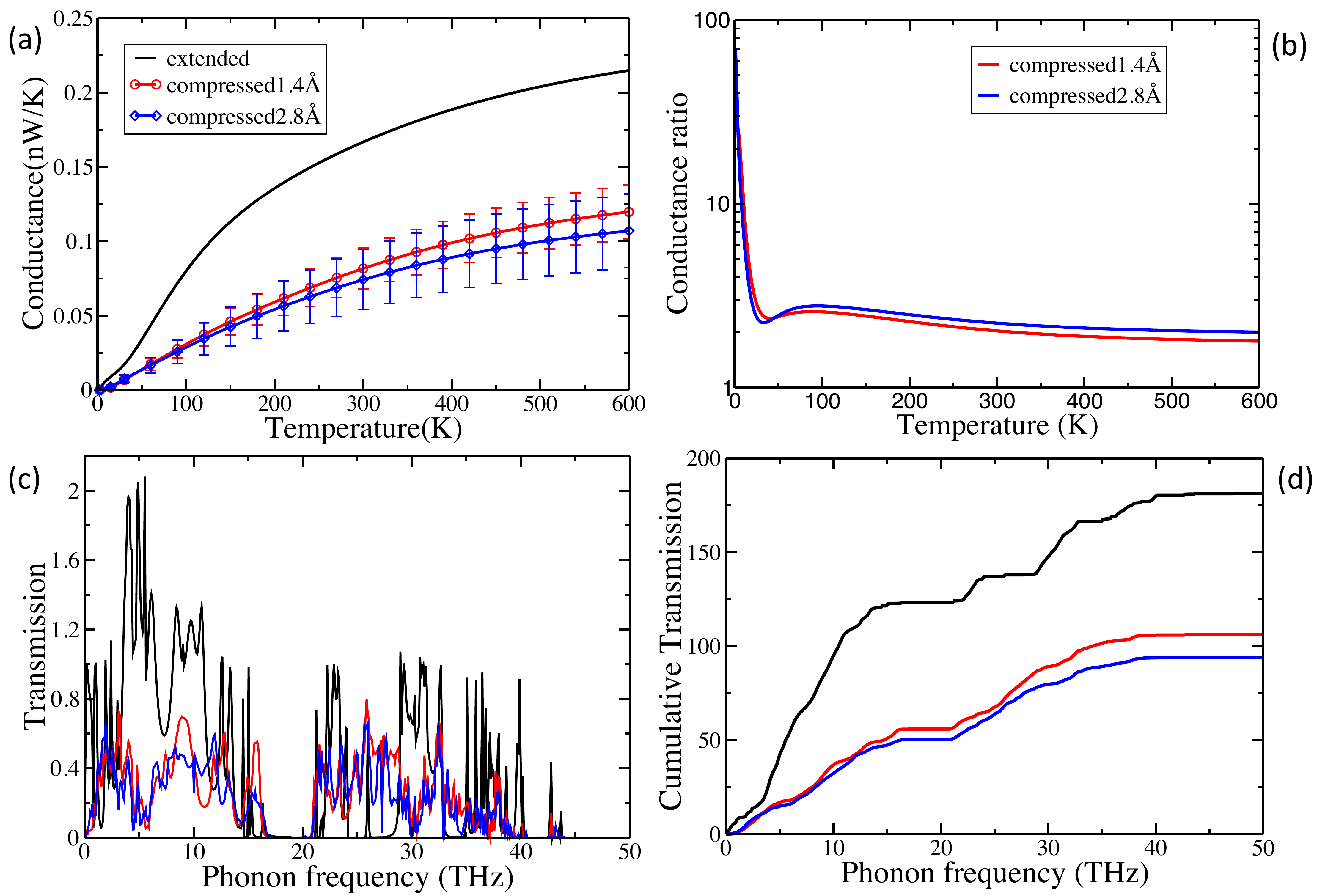}
  \caption{ \textcolor{black}{(a) Thermal conductance as a function of temperature for the extended junction and average conductance for the compressed junctions (with compression ratios of 0.11 corresponding to 1.4 \AA\ and 0.22 corresponding to 2.8 \AA\ ).The error bars give the standard deviation of the conductance. (b) Conductance ratio, defined as the ratio of $G_{extended}/G_{compressed}$. (c) Phonon transmission as a function of frequencies for extended junction and average transmission for the compressed junction. (d) Frequency dependent cumulative phonon transmission.}}
  \label{conduct}
\end{figure}

\begin{figure}[H]
\centering
  \includegraphics[width=1.0\linewidth]{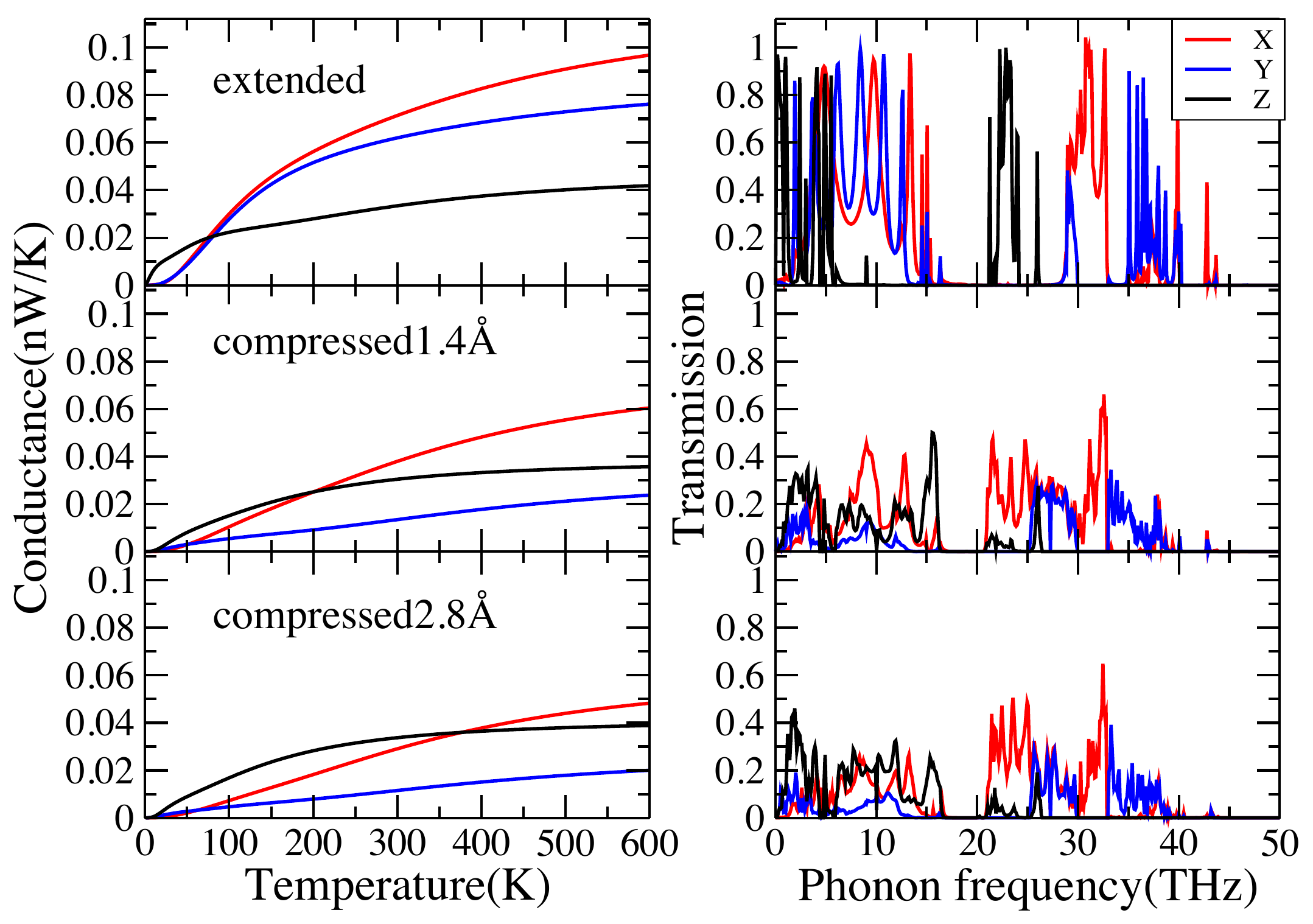}
  \caption{Average phonon transmission and thermal conductance selected according to the polarization of the incoming phonon channels of the graphene leads, for the extended (top panels) and  \textcolor{black}{the compressed 1.4 \AA\ (compression ratio 0.11) and 2.8 \AA\ (compression ratio 0.22) junctions (middle and bottom panels)}. X, Y and Z components correspond to longitudinal, in-plane transverse and flexural modes.}
  \label{xyz}
\end{figure}

Figure~\ref{conduct}(a) shows the thermal conductance $G$ as a function of temperature for the extended molecular junction and the average conductance for two molecular junctions with different compression ratios. The error bars represent the standard deviation of the conductance. The extended molecular junction exhibits relatively high thermal conductance $G_{extended}$ and constitutes an ``on" state. Conversely, the compressed junction has a systematically lower conductance $G_{compressed}$, which is about one half of  $G_{extended}$ at temperature larger than 150 K. This significant reduction of conductance constitutes an ``off" state. 
Here we also note that although the compression ratio of the second junction (0.22) is twice as large as the first one (0.11), the average conductance does not change significantly. Our understanding of this behaviour is the following: when the molecule is compressed by 1.4 \AA, the symmetry is broken and the scattering effect is already strong. Thus, when the junction is further compressed, by 2.8 \AA, the scattering effect is not enhanced.  
We define the \textcolor{black}{conductance ratio} as $G_{extended}/G_{compressed}$ (Fig.~\ref{conduct}(b)). At very low temperature the \textcolor{black}{conductance ratio} is very large, indicating that the lowest frequency acoustic modes, which are the only ones populated at low temperature, are strongly affected by the conformational changes of the alkane chain, but it rapidly drops to about 2.5 at about 50 K, and for $T$ larger than 100 K, it slowly decreases to $\sim 2$.   
Figure~\ref{conduct}(c) shows the phonon transmission functions of the junctions and allows us to interpret the behavior of the thermal conductance and of the \textcolor{black}{conductance ratio} discussed above. The transmission function of the extended junction presents intense peaks at low frequencies indicating strong coupling between the chain and the graphene leads. In turn, the phonon transmission of the compressed junctions decreases dramatically, compared to that of the extended junction, especially in the frequency range from 0 THz to 15 THz. We observe a gap in the transmission spectrum between 15 THz and 22 THz for the extended junction, which was formerly identified as characteristic of the vibrational spectrum of alkane chains~\cite{Nitzan}. Remarkably, this feature persists also when the chain is compressed. 
Above 22 THz, the transmission of compressed junctions is only slightly reduced, compared with that of the extended one. 
 It is also evident that below 1.5 THz, the compressed junctions show almost zero transmission, indicating that bending is very efficient to block extremely long wave length phonons. This is acutally due to the effect on the out-of-plane phonons as shown by the transmission function by polarization (Fig.~\ref{xyz}). This  also explains the very large \textcolor{black}{conductance ratio} at very low temperature.
These features lead to the cumulative transmission function (Fig.~\ref{conduct}(d)), where the slope for the extended junction is almost 3 times larger than that of the compressed one between 0-15 THz, while above 15 THz, the slopes of different cases are very similar. 
The reduced transmission at low frequencies, combined to the almost unchanged transmission at high frequency, finally leads to the temperature dependence of the relative conductances shown in Fig.~\ref{conduct}. Fig.~\ref{conduct}(d) shows that the cumulative transmission of the 1.4 \AA\ and 2.8 \AA\ compressed junctions nearly overlap, resulting in similar thermal conductance. 

To gain further insight into the transmission spectrum, we also study the phonon conductance and transmission by polarization of the incoming channels. Fig.~\ref{xyz} shows the average conductance and transmission contributions from $x$ (longitudinal), $y$ (in-plane transverse), and $z$ (out of plane) channels for compressed junctions, compared to those of the extended one (top panel). 
The thermal conductance of the extended junction is dominated by the $z$ channel below 70 K, but eventually the in-plane $x$ and $y$ channels provide the largest contributions at higher temperatures. 
Bending of the alkane chain largely reduces the conductance through the $x$ and $y$ channels,  while the conductance from out-of-plane modes  remains nearly unchanged. As a consequence the thermal conductance of compressed junctions is mainly reduced by changes in the transmission function of the in-plane modes.
From the phonon transmission by polarization in Fig.~\ref{xyz}(right panel), we find that the average transmission from the longitudinal and the in-plane transverse channels is largely attenuated in the compressed junctions, especially in the low frequency range. 
We observe a wide zero-transmission gap between 8 THz and 22 THz in the transmission function of the out-of-plane (flexural) modes for the extended junction, while this gap is reduced, and more peaks appear in the 8--16 THz range for the compressed junctions. This effect leads to a relatively high conductance from the flexural modes also in the compressed junctions, especially at room temperature and higher. \textcolor{black}{Further reduction of the thermal conductance of the system should therefore focus on reducing the contribution of flexural modes to the conductance of the compressed chain.}
 
In order to identify the origin of the reduced conductance upon compression, we compare the mode localization of an extended molecule with two molecules in the compressed (1.4 \AA\ and 2.8 \AA) junctions. The participation ratio characterizes each mode individually and serves as a useful measure of spatial localization. The participation ratio $p_\lambda$ is defined for each mode $\lambda$ by~\cite{Wooten1993}: 
\begin{equation}
p_{\lambda} =  \frac{1}{N\sum_i (\sum_\alpha \epsilon_{i\alpha,\lambda}^* \epsilon_{i\alpha,\lambda})^2}
\end{equation}
where $\epsilon_{i\alpha,\lambda}$ is the $\alpha$ Cartesian component of the polarization vector of the $\lambda$th mode on the $i$th atom and $N$ is the number of atoms. 
The participation ratio measures the fraction of atoms participating in a mode. If a vibrational mode is delocalized, $p_\lambda$ approaches unity, whereas if a vibrational mode is localized, $p_\lambda$ is very small($\sim 1/N$). 
 \begin{figure}[H]
\centering
 \includegraphics[width=0.8\linewidth]{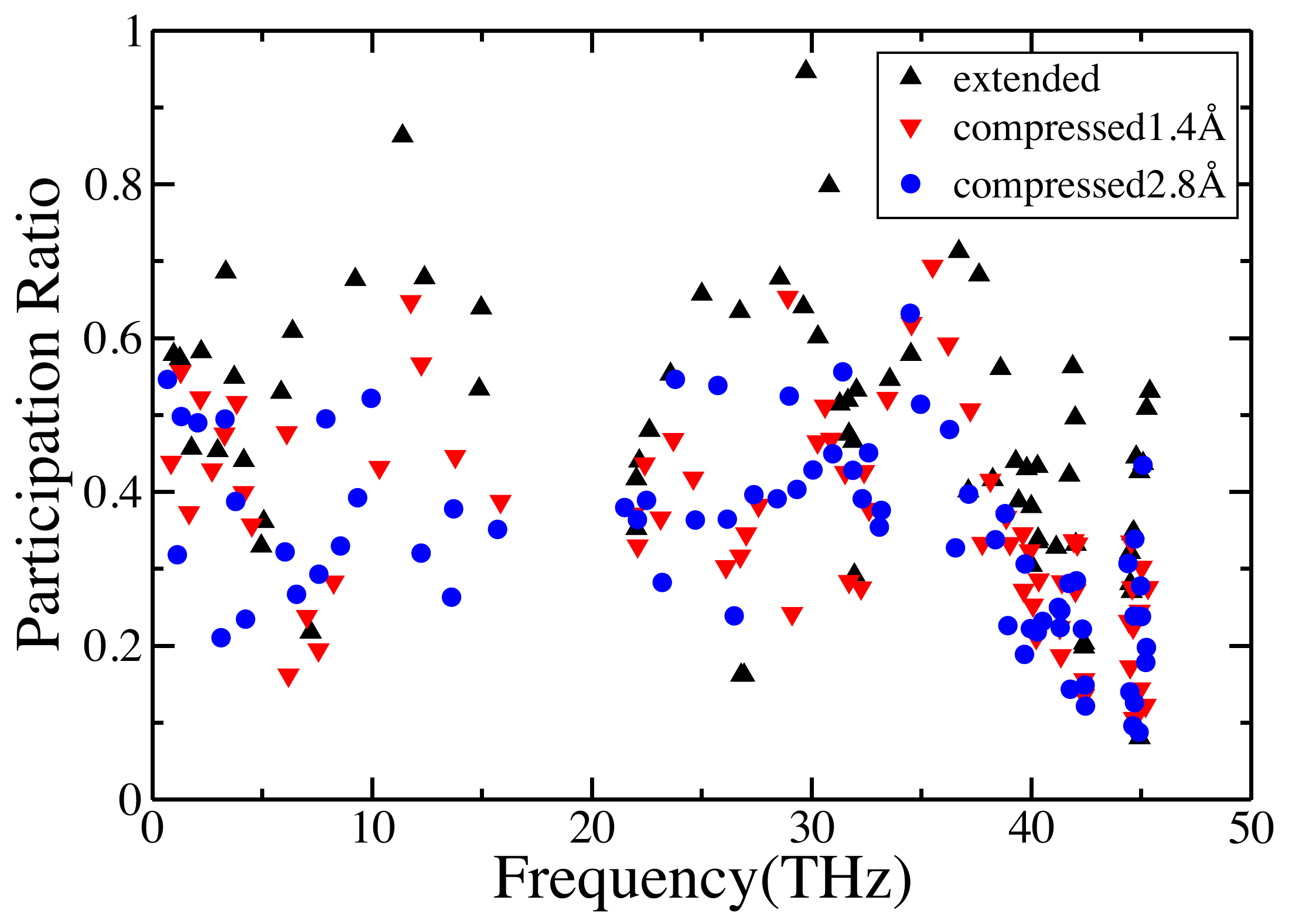}
 \caption{\textcolor{black}{Participation ratio for an extended molecule and two compressed molecules (with compression ratios of 0.11 corresponding to 1.4 \AA\ and 0.22 corresponding to 2.8 \AA\ ).}}
 \label{particip}
\end{figure}

\textcolor{black}{Fig.~\ref{particip} compares the participation ratios for all modes of the extended molecule and two compressed molecules with compression ratios of 0.11 corresponding to 1.4 \AA\ and 0.22 corresponding to 2.8 \AA .}
The participation ratios in the two compressed molecules are generally lower than that of the extended chain over the whole spectrum of frequencies, indicating that modes undergo partial localization upon compression, which accounts for a significant decrease of phonon transmission. Localized modes are indeed much less effective heat carriers~\cite{Wooten1993}.  

\begin{figure}[H]
\centering
  \includegraphics[width=0.75\linewidth]{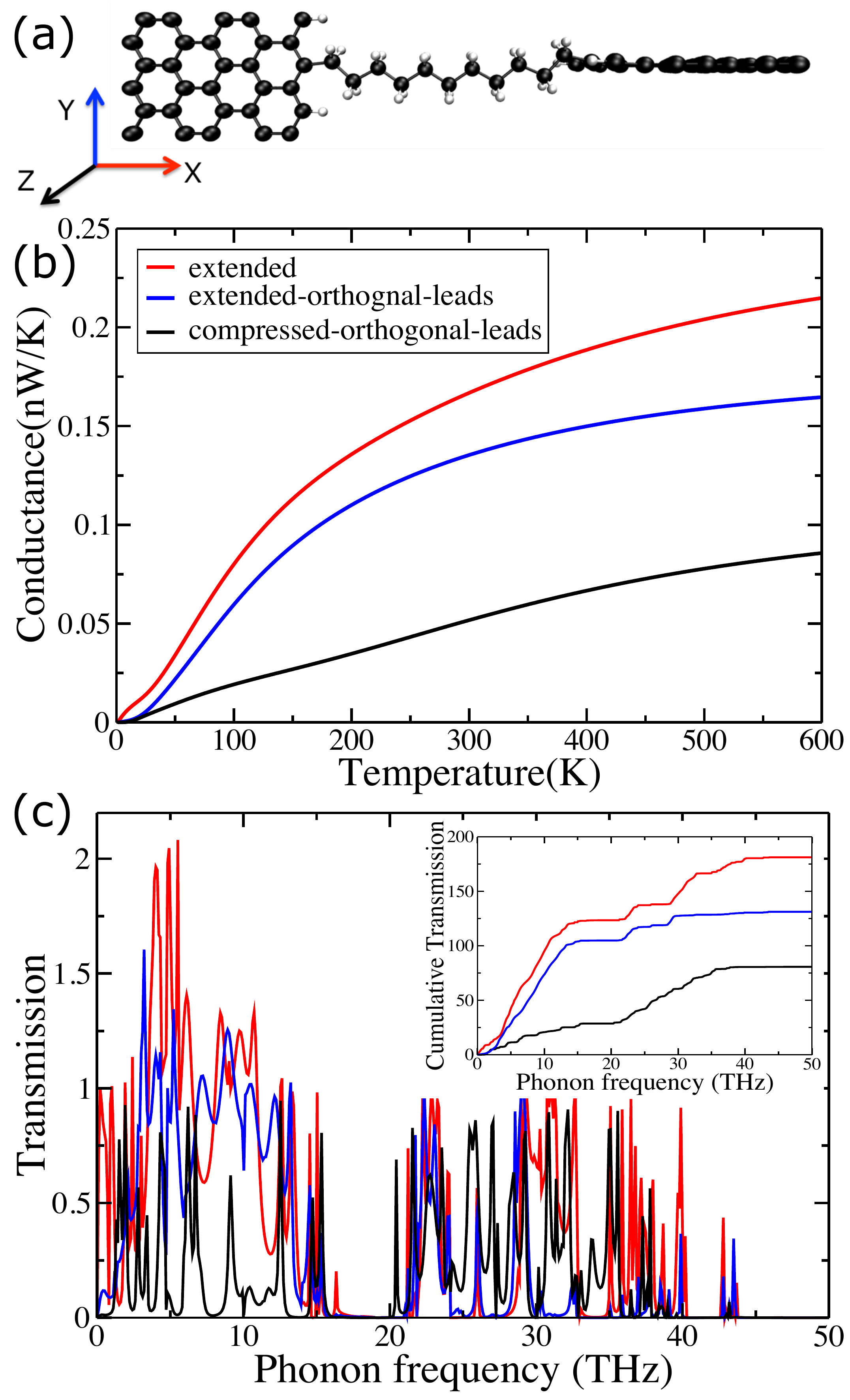}
  \caption{(a) Thermal conductance as a function of temperature and (b) phonon transmission as a function of frequency for the extended junction with co-planar and orthogonal leads and for a single configuration of the compressed (with compression ratios of 0.22 corresponding to 2.8 \AA) junction with orthogonal leads. Inset: cumulative transmission for these three junctions.}
  \label{orthogonal}
\end{figure}

The junction considered so far is an ideal case and one significant assumption is that of perfectly co-planar leads. To test whether the thermal switching mechanism would persist upon misalignment of the leads, we rotate one lead by 90$^{\circ}$ and compute the thermal conductance with the same approach described above. An extended molecule is covalently connected between two orthogonal graphene leads, as shown in the inset of Fig.~\ref{orthogonal}(a). 
Fig.~\ref{orthogonal}(b) shows that while there is a difference, the thermal conductance of the orthogonal junction does not change very much at low temperatures and only decreases by 23\% at 600 K, compared with the co-planar extended junction. The phonon transmission of the orthogonal junction differs slightly from the co-planar extended junction at low frequencies and decreases significantly only above 30 THz, as shown in Fig.~\ref{orthogonal}(c). 
The inset of Fig.~\ref{orthogonal}(c) also shows that the integrated transmission functions are very similar for the extended junction with co-planar and orthogonal leads  
below 30 THz, while differences appear between 30 and 40 THz. 
As the temperature increases, these high-frequency vibrational states get populated and influence the overall conductance more significantly, making the conductances  differ more. 

The thermal conductance and the transmission function for a representative structure of the most populated cluster of configurations of a compressed chain with orthogonal leads are also shown in Fig.~\ref{orthogonal}~(b) and (c). 
{As in the case of co-planar leads, the thermal conductance of the compressed junction with orthogonal leads is about one half of that of the extended junction with orthogonal leads at room temperature.}
The transmission function and its cumulative integral (Fig.~\ref{orthogonal}c) show that the main differences between the extended and the compressed junction stem from modes with frequency lower than 15 THz, i.e. from the acoustic modes below the transmission gap. 
The similar offset in thermal conductance in both the extended and compressed configuration, occurring upon misalignment of the leads, produces a slight enhancement of the \textcolor{black}{conductance ratio}, thus suggesting that the switching mechanism is solid and would be preserved, if not even enhanced, also upon random misalignment of the device. 

\section{Conclusions}

In summary, we found that the thermal conductance of a molecular junction made of a short alkane chain clamped between graphene leads can be controlled by mechanically shifting the leads from a configuration in which the molecule is extended to one in which the molecule is compressed, i.e. presents gauche defects.
The average conductance of the compressed junction drops by 50\% above 150K, compared to that of extended junction, regardless the compression ratio. 
This effect opens the possibility of turning on and off the heat current through the molecular junction, \textcolor{black}{as in a mechanically actuated modulation.} 
The total transmission of the compressed junction decreases dramatically between 0 THz and 15 THz, which is the main reason for the reduction of the thermal conductance.  
The modes of the alkane chain undergo localization upon compression leading to the observed decrease of phonon transmission. 
The reduced thermal conductance in the compressed junction mainly stems from a large suppression of the transmission coefficients of the longitudinal and the in-plane transverse channels of the leads.
Calculations of thermal transport across a molecular junction in which the graphene leads are orthogonal to one another show the same effects as in the one with co-planar leads,  proving that the switching mechanism would persist even if the two graphene leads are not perfectly aligned. 

Our study sheds light on fundamental mechanisms of phonon transport through molecular junctions tuned by physical perturbation and suggests the possibility of making mechanically actuated thermal modulation at the molecular scale. 
\textcolor{black}{Furthermore, since the phononic thermal conductance is a limiting factor in increasing the thermoelectric efficiency in molecular junctions, our work suggests a general way to reduce G, which may improve the thermoelectric figure of merit, provided that the power factor is not reduced upon mechanical compression. However, the effect of compression on the electron transport properties of a molecular junction remains an open issue.}

\begin{acknowledgement}

This work was supported by the funding from the European Research Council under the European Union's Seventh Framework Program (FP7/2007- 2013)/ERC Grant Agreement No. 258806 and The Danish Council for Independent Research Natural Sciences.
Calculations have been performed at the Rechenzentrum Garching of the Max Planck Society.

\end{acknowledgement}

\bibliography{Molecular-junctions-JPC}

\end{document}